\documentclass[]{article}

 \usepackage[]{graphicx}
 \usepackage{epsfig}
 \usepackage{amsmath, amstext, amsfonts}
 \usepackage{amssymb}
 \usepackage{multicol}
 \usepackage[latin1]{inputenc}

\usepackage{anysize} 
\marginsize{2cm}{2.5cm}{1.5cm}{2cm}

 \title{A note on Verhulst's logistic equation and related logistic maps}

 \author{M. Ranferi Guti\'errez$^1$, M.A. Reyes$^1$, and H.C. Rosu$^2$\footnote{E-mail: hcr@ipicyt.edu.mx}\\
 {\small \it $^1$ Depto de F\'{\i}sica, Universidad de Guanajuato}, {\small \it Apdo. Postal E143, 37150 Le\'on, Gto., Mexico}\\
 {\small \it $^2$ IPICyT, Instituto Potosino de Investigacion Cientifica y Tecnologica},\\
{\small \it Apdo Postal 3-74 Tangamanga, 78231 San Luis Potos\'{\i}, Mexico}}

 \date{J. Phys. A 43 (2010) 205204 (5pp)\\
 {\tiny doi:10.1088/1751-8113/43/20/205204}\\ {\tiny arXiv:0910.1560v2 [math-ph]}\\ {\tiny Logistic-eq-map.tex}}

\begin{document}

 \maketitle
 \begin{center}{\sc 
 \footnotesize \vskip 4ex
}
 \end{center} 

\medskip

\noindent PACS {\tt 02.60.Cb} - Numerical simulation; solution of equations\\
PACS {\tt 05.45.Ac} - Low-dimensional chaos \\
PACS {\tt 05.45.Pq} - Numerical simulations of chaotic systems \\

\bigskip

\begin{abstract}
  \footnotesize
   \noindent
   We consider the Verhulst logistic equation and a couple of forms of the corresponding logistic maps. For the case
of the logistic equation we show that using the general Riccati solution only
changes the initial conditions of the equation. Next, we consider two forms of corresponding
logistic maps reporting the following results. For the map $x_{n+1} = rx_n(1 - x_n)$ we propose a new way to write the
solution for $r = -2$ which allows better precision of the iterative terms, while for
the map $x_{n+1}-x_n = rx_n(1 - x_{n+1})$ we show that it behaves identically to the logistic
equation from the standpoint of the general Riccati solution, which is also provided herein for any value of the parameter $r$.
\\

 \end{abstract}


{\bf Introduction}. -
Verhulst's equation, first discussed by P.-F. Verhulst in 1845 and 1847 and rediscovered during the 1920s, is also known more recently as the logistic equation
\begin{equation} \label{ranf-1}
\dot{x}=rx(1-x)~.
\end{equation}
It is a Riccati equation of constant coefficients that has been applied in very different areas of science, like biology,
demography, economy, chemistry, and probability and statistics \cite{May76}. The solution to this
equation can be obtained after multiplying by the integrating factor $e^{rt}/x^2$ that allows to write it as $\frac{d}{dt}\left(\frac{e^{rt}}{x}\right)=re^{rt}$ which is readily integrated leading to the solution
\begin{equation} \label{ranf-2}
x_1(t) =\frac{1}{1+(x_{0}^{-1}-1)e^{-rt}}~,
\end{equation}
where $x_0=x(0)$. After the influential results of A.J. Lotka in 1925, eq.~(1) is considered as the basic equation for understanding population growth. The discrete equations derived from it are known as logistic maps and have been thoroughly investigated in the literature \cite{Japan}. The most renowned of these maps is
\begin{equation} \label{ranf-3}
x_{n+1} = rx_n(1 - x_{n})
\end{equation}
since it is known that for $r\geq 4$ it describes chaos of discrete type for almost all initial values \cite{elaydi}, and can therefore be used as a random number generator. There are very few known solutions to this equation, namely for
$r = -2,\, 2, \,4$, which might be the only possible solutions in closed form.
On the other hand, a different discrete form which keeps closer resemblance to eq.~(1) is the following logistic map
\begin{equation} \label{ranf-4}
x_{n+1}-x_n = rx_n(1 - x_{n+1})
\end{equation}
since its solution is
\begin{equation} \label{ranf-5}
x_{n,1} =\frac{1}{1+(x_{0}^{-1}-1)(1+r)^{-n}}~,
\end{equation}
in analogy to solution (2) of eq.~(1).
In this brief work we shall deal with exact solutions of eqs.~(1), (3), and (4).

\medskip

\begin{figure}
\centering
\includegraphics[height=4.4622in,width=5.9055in] {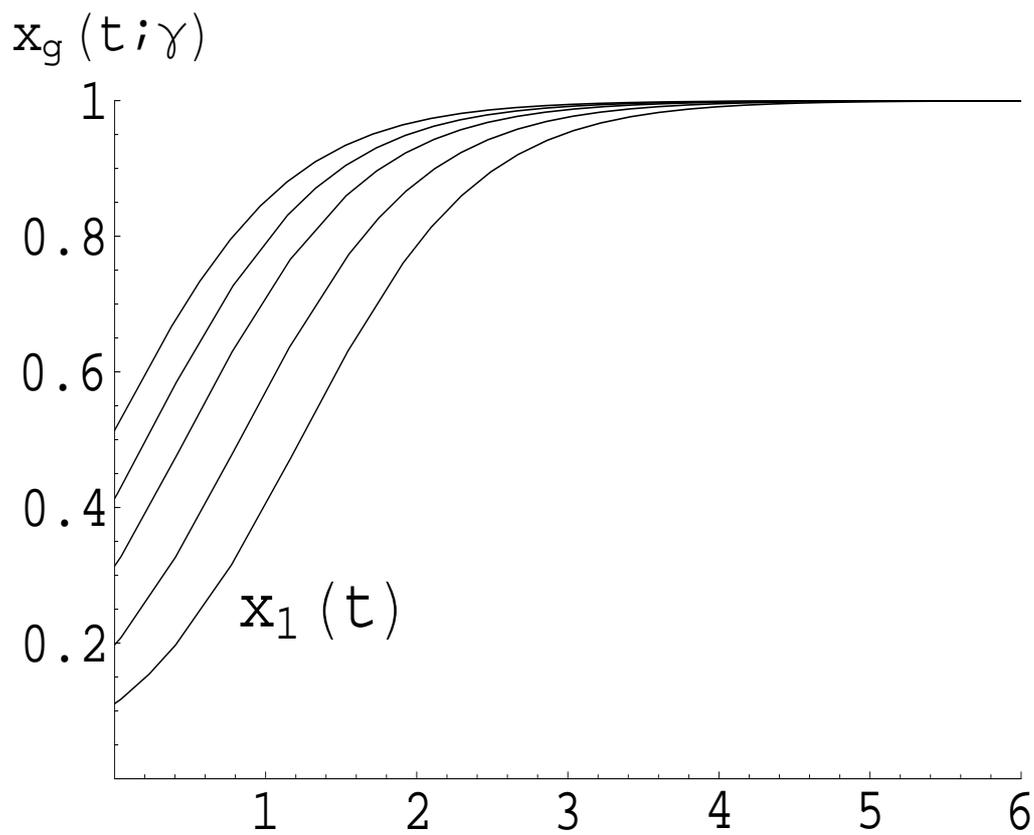}
\caption{Solutions of the logistic equation for the initial condition $x_0=0.11$ and $r=1.7$ and increasing values of the parameter $\gamma=0.14, 0.15, 0.17, 0.25$ from top to $x_1(t)$, respectively.}
\label{figure1}
\end{figure}

{\bf Logistic equation}. -
We first consider the logistic equation (1). Since it is a Riccati equation, and we already
know one particular solution to this equation, we can try finding the general solution in the
form $x_g(t) = x_1(t) + v(t)$, where $x_1(t)$ is given in eq.~(2). With the change $v(t) = 1/y(t)$,
$y$ has to satisfy the linear equation
\begin{equation} \label{ranf-6}
\dot{y} + r (1 - 2x_1) y = r~.
\end{equation}
Integrating this equation using $y(0) =\gamma$, we find that
\begin{equation} \label{ranf-7}
x_g(t) = x_1(t)\left(
1+\frac{1}{\gamma \left(e^{rt}+(x_{0}^{-1}-1)\right)-1}\right)~.
\end{equation}
The latter equation can also be written in the following form
\begin{equation} \label{ranf-8}
x_g(t) = \frac{1}{1+\left(\frac{\gamma - x_0}{\gamma x_0}-1\right)e^{-rt}}~,
\end{equation}
which is just the solution (2) with the initial condition $x_{\gamma}=\frac{\gamma x_0}{\gamma -x _0}$.
Therefore, the parameter $\gamma$ has to be in the range $\frac{x_0}{1-x_0}<\gamma < \infty$. In Fig.~(\ref{figure1}) plots of the solution
(\ref{ranf-8}) are displayed for several values of the parameter $\gamma$.

\medskip

{\bf Logistic maps}. - Now let us deal with the aforementioned logistic maps:

\medskip

(a) $ x_{n+1} = r x_n(1 - x_n)$\\
For this map, Wolfram postulated that there are only three known solutions, for $r = -2,\, 2\, {\rm and}\, 4$ that can be derived from the form \cite{W}
\begin{equation} \label{ranf-9}
x_n=\frac{1}{2}\left[1-f\left(r^nf^{-1}(1-2x_0)\right)\right]~.
\end{equation}
The three known solutions and the corresponding $f$ are given in Table 1.

\begin{center}
\vspace*{3mm}
\begin{tabular}{|r|c|c|}
\hline
$r$ & $f(x)$ & Solution \\
\hline
-2 & $2 \cos\left(\frac{1}{3}\left(\pi-\sqrt{3} \, x\right)\right)$ &
$\frac{1}{2}-\cos\left(\frac{1}{3}
\left[\pi-(-2)^n\left(\pi-3\cos^{-1}\left(\frac{1}{2}-x_0\right)\right)\right]\right)$ \\
\hline
2 & $e^x$ & $\frac{1}{2}\left[1-\exp\left(2^n \ln(1-2\, x_0)\right)\right]$ \\
\hline
4 & $\cos x$ & $\frac{1}{2}\left[1-\cos\left(2^n \cos^{-1}(1-2\, x_0)\right)\right]$   \\
\hline
\end{tabular}
\vspace*{3mm}
\\
{Table 1. The three exact solutions derived from the formula (\ref{ranf-9}).}
\vspace*{3mm}
\end{center}

\bigskip

In fact, the solution for $r = 4$ was first introduced by Ulam and von Neumann \cite{UvN}:
proposing $x_n = \sin^2(\pi z(n))$ we get that $z(n) = 2^n z(0)$, leading to the solution in Table
1.

Actually, we shall show here that the three solutions in Table 1 can easily be obtained
in the following way. Beginning with eq.~(3) written as $x_{n+1} = -r(x^{2}_{n}-x_n)$, let us complete
squares and define $y_n = x_n - 1/2$. Then $y_n$ obeys the equation
\begin{equation}
y_{n+1}=-r\, y_n^2+\left( \frac{r}{4}-\frac{1}{2} \right)
\end{equation}

\noindent
{\boldmath $r=2$}. We can see that if $r=2$, $y_{n+1}=-2\, y_n^2$ and then
\begin{equation}
x_n = \frac{1}{2} \left( 1-(1-2\, x_0)^{2^n} \right)
\end{equation}
which is the solution in Table 1.

If $r\neq 2$, we can use the Ulam-von Neumann ansatz and propose
$y_{n}=a\, \cos z_n$, and $z_{n+1}=2\, z_n$.  To complete the trigonometric identity $\cos2\theta=2\, \cos^2\theta-1$ we must have that $a=-2/r$ and that $r$ satisfies $r^2-2r-8=0$, whose solutions are $r=4$ and $r=-2$.

\medskip

\noindent
{\boldmath $r=4$}. In this case, $x_n$ is given by
\begin{equation}
x_n = \frac{1}{2} \left( 1-\cos(2^n \, \cos^{-1}(1-2\, x_0)) \right)
\end{equation}

\noindent
{\boldmath $r=-2$}. In this case, $x_n$ is given by
\begin{equation}
x_n = \frac{1}{2} +\cos(2^n \, \cos^{-1}(1-2\, x_0))~.
\label{req-2}
\end{equation}

\bigskip

One can see that the latter result appears to be different from that in Table 1 for $r = -2$.
However, one can show that these two solutions can be transformed from one to the other
using trigonometric identities. Nonetheless, it is well known that the logistic map (3) is
very dependent on small fluctuations and on numerical exactness of the computing device
used to calculate it. Therefore eq.~(13), which is a simpler form of the solution, can be
used more accurately than the solution given in Table 1, as can be seen in fig.~(2), where
the three alternatives discussed herein are displayed.

\begin{figure}
\centering
\includegraphics[height=4.4622in,width=5.9055in] {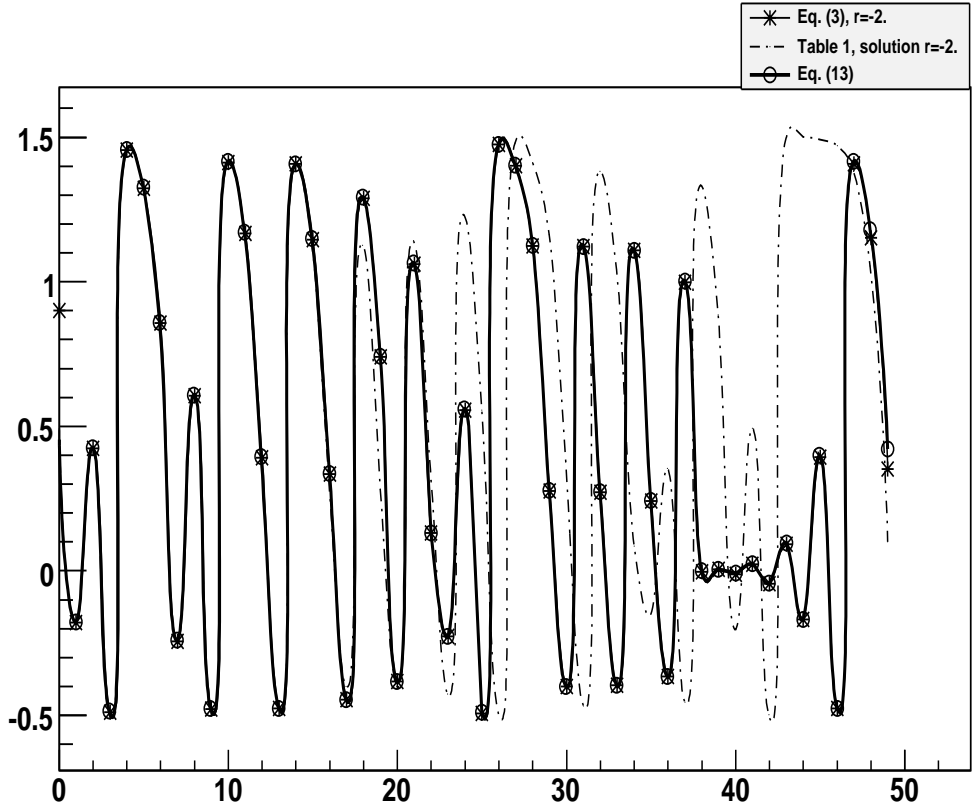}
\caption{ Three plots of the logistic map (3) for $r=-2$ and the same initial condition $x_0=0.9$. With stars we plot the numerical iterations of the map; the line-dot curve corresponds to the values calculated with the solutions in the form given in Table 1 and the continuous curve shows the values from solution (13), which behaves just like the iterative points.}
\label{figure2}
\end{figure}

\medskip

(b) $x_{n+1}-x_n = r x_n(1 - x_{n+1})$\\
Contrary to the previous logistic map, this one has the exact solution (5). Even more, it
is possible to develop the general Riccati solution for this difference equation, as has been
done before for the three site master equation \cite{ranf-2}. By defining $x_{n,1}$ as given by solution (5),
the general Riccati solution turns out to be
\begin{equation} \label{ranf-10}
x_{n,g}=x_{n,1}+\frac{\prod _{k=0}^{n-1}g_k^{-1}}{\gamma +\sum _{k=0}^{n-1}\left(\prod_{j=0}^{k}g_{j}^{-1}\right)h_k}
\end{equation}
where
\[
g_n=\frac{r\, x_{n,1}+1}{r(1-x_{n+1,1})+1}
\ \ \ \ \mbox{and} \ \ \ \ h_n=\frac{r}{r(1-x_{n+1,1})+1}~.
\]
Even when this solution is difficult to interpret as the different-initial-condition form of the difference equation (\ref{ranf-4}), it behaves just like in the continuum case, as can be seen in Fig.~(3), where different values of the parameter $\gamma$ lead to different initial conditions.

\medskip

\begin{figure}
\centering
\includegraphics[height=4.4622in,width=5.9055in] {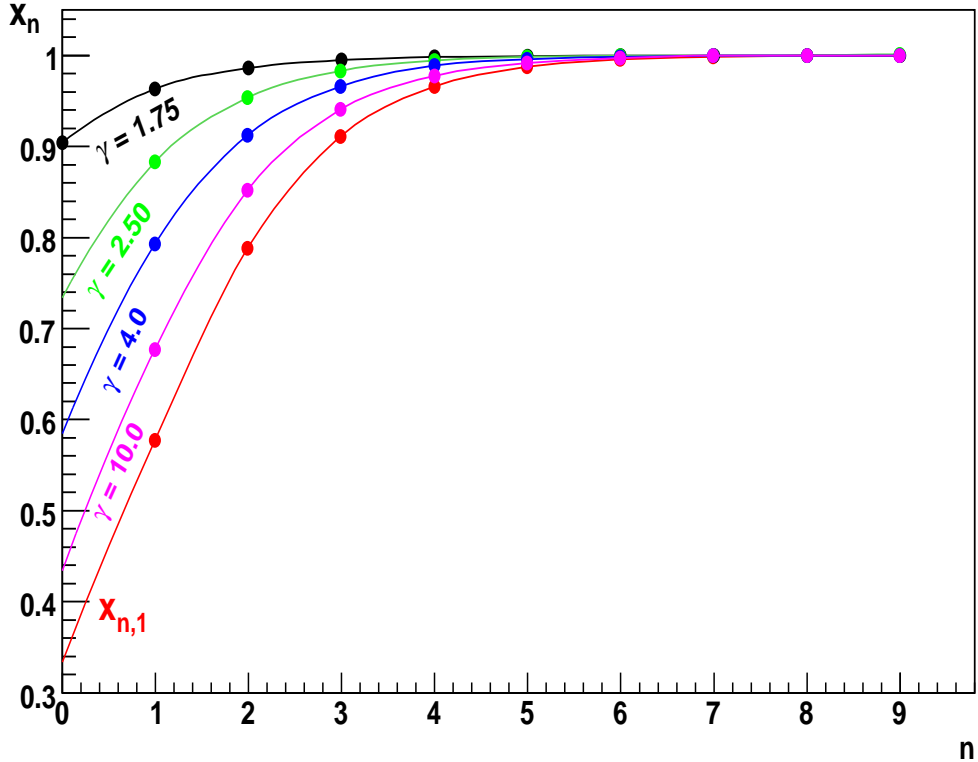}
\caption{ The general Riccati solution for the map (4) with $x_0=0.333$ and $r=1.73$ for different values of the parameter $\gamma$. These solutions behave exactly as the solutions of the logistic equation (the continuous case).
}
\label{figure3}
\end{figure}

{\bf Conclusion}. -
In this short note we have dealt with exact solutions of the logistic equation and logistic maps.
After presenting the general Riccati solution for the logistic equation, we introduce a simpler form of the solution of the standard logistic map (3) which is more accurate than the solution cited in the literature. We also show that the slightly modified logistic map (4) is the
closest difference representation of the logistic equation from the point of view of its general Riccati solution. Moreover, in the latter case, the general solution (14) that we provide here is valid for any value of the parameter $r$ whereas in the first case the analytic solution is restricted to only three values so far. For a different argument in favor of the form (4), the reader is directed to the recent work of Takenouchi and Ota \cite{TO}.


\begin{center} *** \end{center}

We acknowledge support from CONACYT of Mexico, through a scholarship for M. Ranferi Guti\'errez.

\bigskip


\end{document}